\documentclass[aps,groupedaddress,superscriptaddress,amsmath,amssymb,twocolumn,pra]{revtex4-1}
\usepackage{bm}
\usepackage[dvips]{graphicx}
\usepackage{wrapfig,subfigure}
\usepackage{stmaryrd}
\usepackage{color}
\usepackage[normalem]{ulem}
\usepackage{enumerate}
\usepackage{epstopdf}

\def\HR{H_{\rm RWA}}

\def\rb{{\bf r}}

\begin{document}
\author{Mark Dykman}
\affiliation{Department of Physics and Astronomy, Michigan State University, East Lansing, MI 48824, USA}
\author{Gianluca Rastelli}
\affiliation{Fachbereich Physik, Universit\"at Konstanz, D-78457 Konstanz, Germany}
\author{M. L. Roukes}
\affiliation{Department of Physics, California Institute of Technology, Pasadena, CA 91125, USA}
\author{Eva M. Weg}
\affiliation{Fachbereich Physik, Universit\"at Konstanz, D-78457 Konstanz, Germany}

\title{Resonantly induced friction in driven nanomechanical systems}

\date{\today}
\begin{abstract}
We propose a new mechanism of friction in  resonantly driven vibrational systems. The form of the friction force follows from the time- and spatial-symmetry arguments. We consider a microscopic mechanism of this resonant force in nanomechanical systems. The friction can be negative, leading to an instability of forced vibrations of a nanoresonator and the onset of self-sustained oscillations in the rotating frame. 
\end{abstract}
\maketitle


The physics of  friction keeps attracting attention in diverse fields and at different spatial scales, from cold atoms to electrons on helium to locomotion of devices and animals \cite{Vanossi2013,Bylinskii2015,Rees2016,Crassous2017,Tian2018}. 
An important type of systems where friction plays a critical role and where it has been studied in depth, both theoretically and experimentally, are vibrational systems. The simplest form of friction in these (and many other) systems is viscous friction.
For a vibrational mode with coordinate $q$, the viscous friction force is $\propto \dot q$. It describes a large number of experiments on various kinds of vibrational systems, nano- and micromechanical modes and electromagnetic cavity modes being examples of the particular recent interest \cite{Walls2008,Schmid2016}. 

In vibrational systems, viscous friction is often called linear friction, to distinguish it from nonlinear friction, which nonlinearly depends on $q$ and $\dot q$. 
Phenomenologically, the simplest nonlinear friction force is $\propto q^2\dot q$ (the van der Pol form \cite{Pol1926}) or $\propto \dot q^3$ (the Rayleigh form \cite{Rayleigh1894}). Both these forms of the force are particularly important  for weakly damped systems. This is because in such systems the vibrations are nearly sinusoidal, whereas both forces have resonant components which oscillate at the mode frequency. Moreover, both forces lead to the same long-term dynamics of a weakly damped mode and in this sense are indistinguishable \cite{Lifshitz2008,Atalaya2016}. 

External driving of vibrational modes can modify their dissipation. The change has been well understood for a periodic driving tuned sufficiently far away from the mode eigenfrequency. Such driving can open new decay channels where transitions between the energy levels of the mode are accompanied by absorption or emission of excitations of the thermal reservoir and a drive quantum $\hbar\omega_F$, with $\omega_F$ being the drive frequency \cite{Dykman1978}. This can lead to both linear \cite{Aspelmeyer2014a,Aspelmeyer2014} and nonlinear friction \cite{Leghtas2015,Dong2018}.  It has been also found that, in microwave cavities and nanomechanical systems, resonant driving can reduce linear friction by slowing down energy transfer from the vibrational mode to two-level systems due to their saturation \cite{Gao2007,Gao2008,Singh2016}. 

In this paper we consider nonlinear friction induced by resonant driving, which  significantly  differs from other forms of friction. We show that, in nanomechanical systems, the proposed friction can become important already for a moderately strong drive and can radically modify the response to the drive, including the onset of slow oscillations of the amplitude and phase of the driven mode with the increasing drive \footnote{A loss of stability of forced vibrations with the increasing resonant driving was recently observed in nanomechanical resonators by E. Weig et al.}. 

Phenomenologically, a mode with inversion symmetry driven by a force $F(t)=F\cos\omega_F t$ can experience a resonant induced friction force (RIFF) of the form
\begin{align}
\label{eq:phenomenological}
f_{\rm RIFF} = \eta_{\rm RIFF} F(t)q\dot q.
\end{align}
Such force has the right spatial symmetry, as it changes sign on spatial inversion ($q\to -q$ and $F\to -F$), and is dissipative, as it changes sign on time inversion $t\to -t$.
The driving frequency $\omega_F$ is assumed to be close to the mode eigenfrequency $\omega_0$, so that the force $f_{\rm RIFF}$ has a resonant component. 
 Parameter $\eta_{\rm RIFF}$ is the friction coefficient, which is undetermined in the phenomenological theory. Not only the magnitude, but also the sign of $\eta_{\rm RIFF}$ are not determined, as the very onset of the force $f_{\rm RIFF}$ is a nonequilibrium phenomenon. 

The form of the RIFF reminds the form of the van der Pol friction force, except that $q^2$ is replaced by $F(t)q$. In some sense, the force $F(t)$ is ``smaller'' than the displacement $q$ near resonance: this is the well-known effect that a small resonant force leads to large vibration amplitude for weak damping. Therefore $f_{\rm RIFF}$ can be significant if there is a mechanism that compensates the relative smallness of $F(t)$. 

For nanomechanical resonators, a simple microscopic mechanism of the RIFF is heating. The absorbed power $F(t)\dot q$ leads to a temperature change $\delta T$, which can be relatively large due to the small thermal capacity of a nanoresonator [generally, the temperature change depends on the coordinates in the resonator, see Appendix]. In turn, the temperature change modifies the resonator eigenfrequency $\omega_0$, for example, due to thermal expansion, cf. \cite{Barton2012,Chien2018}. To the lowest order in $\delta T$, the eigenfrequency change is $\delta\omega_0=-\lambda_\omega \delta T$ with the coefficient $\lambda_\omega$ depending on the spatial structure of the mode and the temperature field.

In many cases, the relaxation time of the temperature in the resonator is much longer then the vibration period $t{}_F =2\pi/\omega_F$. Then the temperature change is proportional to the period-averaged power,
\[\delta T(t) =\lambda_T[F(t)\dot q(t)]_{\rm av} \equiv \lambda_Tt{}_F^{-1}\int_t^{t+t{}_F} dt' F(t')\dot q(t')\]
(in fact, $\delta T$ is spatially nonuniform, see Appendix).
As a result, the restoring force $-m\omega_0^2q$ is incremented by $ f_T$,
\begin{align}
\label{eq:T_force}
f_T (t)= 2m\omega_0 \lambda_\omega \lambda_T [F(t)\dot q(t)]_{\rm av}\,q(t)
\end{align}
The force $f_T(t)$ is a specific form of the RIFF. The thermal mechanism is not the only RIFF mechanism, but it is often important, and moreover, the ratio of the conventional nonlinear friction to the RIFF contains a small parameter, see Appendix.

We now consider the dynamics of a driven nanoresonator in the presence of RIFF.  Nanoresonators are often well described by the Duffing model, which takes into account a quartic nonlinearity \cite{Lifshitz2008}. The Hamiltonian of the Duffing oscillator in the absence of coupling to the thermal reservoir is 
\begin{align}
\label{eq:Hamiltonian_lab}
H_0=\frac{1}{2}(p^2 + \omega_0^2q^2) + \frac{1}{4}\gamma q^4 - qF\cos\omega_Ft.
\end{align}
Here $p$ is the oscillator momentum. We have set the mass $m=1$. For concreteness, we assume that the Duffing nonlinearity parameter $\gamma$ is positive. The driving is assumed resonant, $|\omega_F-\omega_0|\ll \omega_0$, and comparatively weak, so that $|\gamma|\langle q^2\rangle \ll \omega_0^2$. 

To analyze the behavior on the time scale long compared to $\omega_F^{-1}$, one can change to the rotating frame and introduce slowly varying in time canonically conjugate coordinate $q_{\rm 0}$ and momentum $p_{\rm 0}$ (the analogs of the quadrature operators) 
\[q(t)+i\omega_F^{-1}p(t) = (\omega_F)^{-1/2}(q_{\rm 0}+ip_{\rm 0})\exp(-i\omega_Ft).\]
In the standard rotating wave approximation (RWA), from Eq.~(\ref{eq:Hamiltonian_lab}) we obtain Hamiltonian equations for $q_{\rm 0},p_{\rm 0}$ 
with the time-independent Hamiltonian $H_{\rm RWA}$,
\begin{align}
\label{eq:RWA_Hamiltonian}
&(\dot q_{\rm 0})_H = \partial_{p_{\rm 0}} H_{\rm RWA},\quad (\dot p_{\rm 0})_H = -\partial_{q_{\rm 0}} H_{\rm RWA},\nonumber \\
&H_{\rm RWA}(q_{\rm 0},p_{\rm 0})=-\frac{1}{2}\delta\omega(q_{\rm 0}^2+p_{\rm 0}^2) + \frac{3\gamma}{32\omega_F^2}(q_{\rm 0}^2+p_{\rm 0}^2)^2\nonumber\\ 
&\qquad - Fq_{\rm 0}/2\sqrt\omega_F, \qquad \delta\omega = \omega_F-\omega_0.
\end{align}

It is well-known how to incorporate linear friction into the RWA-equations of motion starting from both a microscopic formulation and the phenomenological friction force $-2\Gamma \dot q$ \cite{Bogolyubov1945,Senitzky1960,Schwinger1961,DK_review84}. An extension  to the RIFF is straightforward. Keeping only smoothly varying terms in the equations for $\dot q_{\rm 0},\dot p_{\rm 0}$, in the case of the heating-induced RIFF (\ref{eq:T_force}) we obtain  the following equations of motion:
\begin{align}
\label{eq:eom}
&\dot q_{\rm 0} = -\Gamma q_{\rm 0} - J_T p_{\rm 0}^2 +\partial_{p_{\rm 0}} H_{\rm RWA},\nonumber\\
&\dot p_{\rm 0} = -\Gamma p_{\rm 0} + J_T q_{\rm 0} p_{\rm 0} -\partial_{q_{\rm 0}}H_{\rm RWA}.
\end{align}
Here $J_T=\omega_F^{1/2}F\lambda_\omega \lambda_T /2$. In Eq.~(\ref{eq:eom})
we have disregarded noise. It is typically weak in weakly damped nanoresonators and leads primarily to small fluctuations about the stable states of forced vibrations and occasional switching between the stable states in the range of bistability, cf.~\cite{DK_review84,Aldridge2005,Chan2012,Moser2013,Defoort2015,Davidovikj2016,Dolleman2018} and references therein; here we do not consider these effects.

The parameter $J_T$ that characterizes the RIFF increases with the driving amplitude $F$; the RIFF also increases with the vibration amplitude $A= [(q_{\rm 0}^2+p_{\rm 0}^2)/\omega_F]^{1/2}$. From Eq.~(\ref{eq:eom}), the effects of the RIFF become pronounced for $|J_TA|\sim\Gamma$ and should be seen already for a moderately strong drive if the decay rate $\Gamma$ due to the linear friction is small. 

If the linear friction and the RIFF can be disregarded, the values $(q_{\rm st},p_{\rm st})$ of $(q_{\rm 0},p_{\rm 0})$ at the stationary states  of forced  vibrations are given by the conditions $\partial_{q_{\rm 0}}H_{\rm RWA}=\partial_{p_{\rm 0}}H_{\rm RWA}=0$, which reduce to equations
\begin{align}
\label{eq:stationary_states}
 \frac{3\gamma}{8\omega_F^2}q_{\rm st}^3 - \delta\omega\, q_{\rm st} =F/2\sqrt\omega_F, \qquad p_{\rm st}=0.
\end{align}
The equation for $q_{\rm st}$ has one real root in the range of $F, \delta\omega$ where the oscillator is monostable in the weak dissipation limit or 3 real roots in the range of bistability. In the latter range, of primary interest for the analysis of the RIFF is the root with the maximal $q_{\rm st}$, and in what follows $q_{\rm st}$ refers to this root. For small $\Gamma$ and $J_T=0$ it corresponds to a stable state of forced vibrations at frequency $\omega_F$, as does also the real root $q_{\rm st}$ in the range of monostability \cite{LL_Mechanics2004}. In the both cases, the considered $(q_{\rm st},p_{\rm st})$ corresponds to the minimum of $H_{\rm RWA}$.

For $J_T>0$ the RIFF can lead to instability of the forced vibrations. Indeed, to the leading order in $\Gamma, J_T$, the sum of the eigenvalues of Eqs.~(\ref{eq:eom}) linearized about the stable state is $-2\Gamma + J_Tq_{\rm st}$. When this sum becomes equal to zero, the system undergoes a supercritical Hopf bifurcation. This means that, for $J_Tq_{\rm st}>2\Gamma$, the state of forced vibrations with constant amplitude and phase becomes unstable. The amplitude and phase oscillate in time, which corresponds to oscillations of the system in the rotating frame about $(q_{\rm st}, p_{\rm st})$.

\begin{figure}
\includegraphics[scale=0.28]{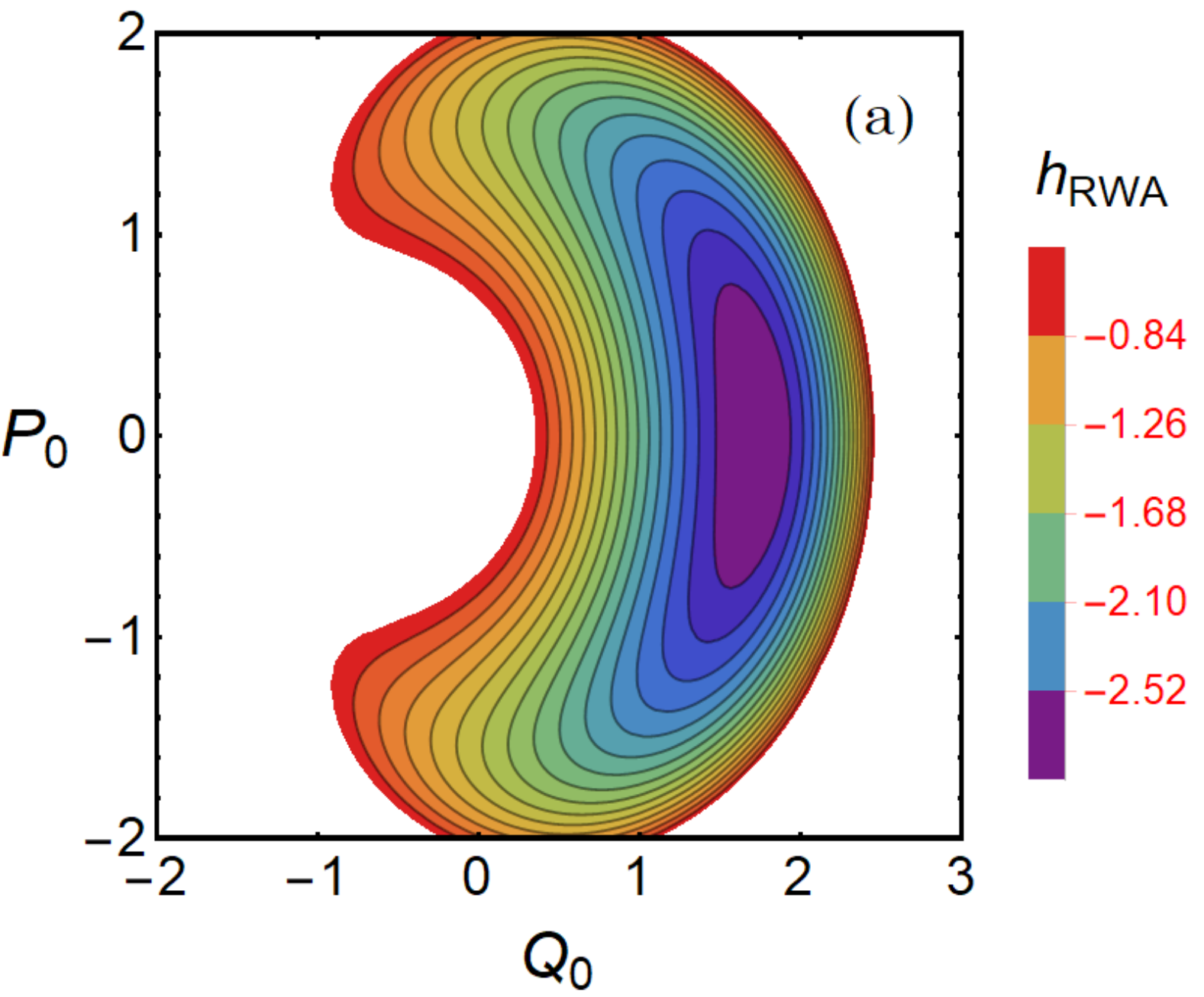}\hfill
\includegraphics[scale=0.3]{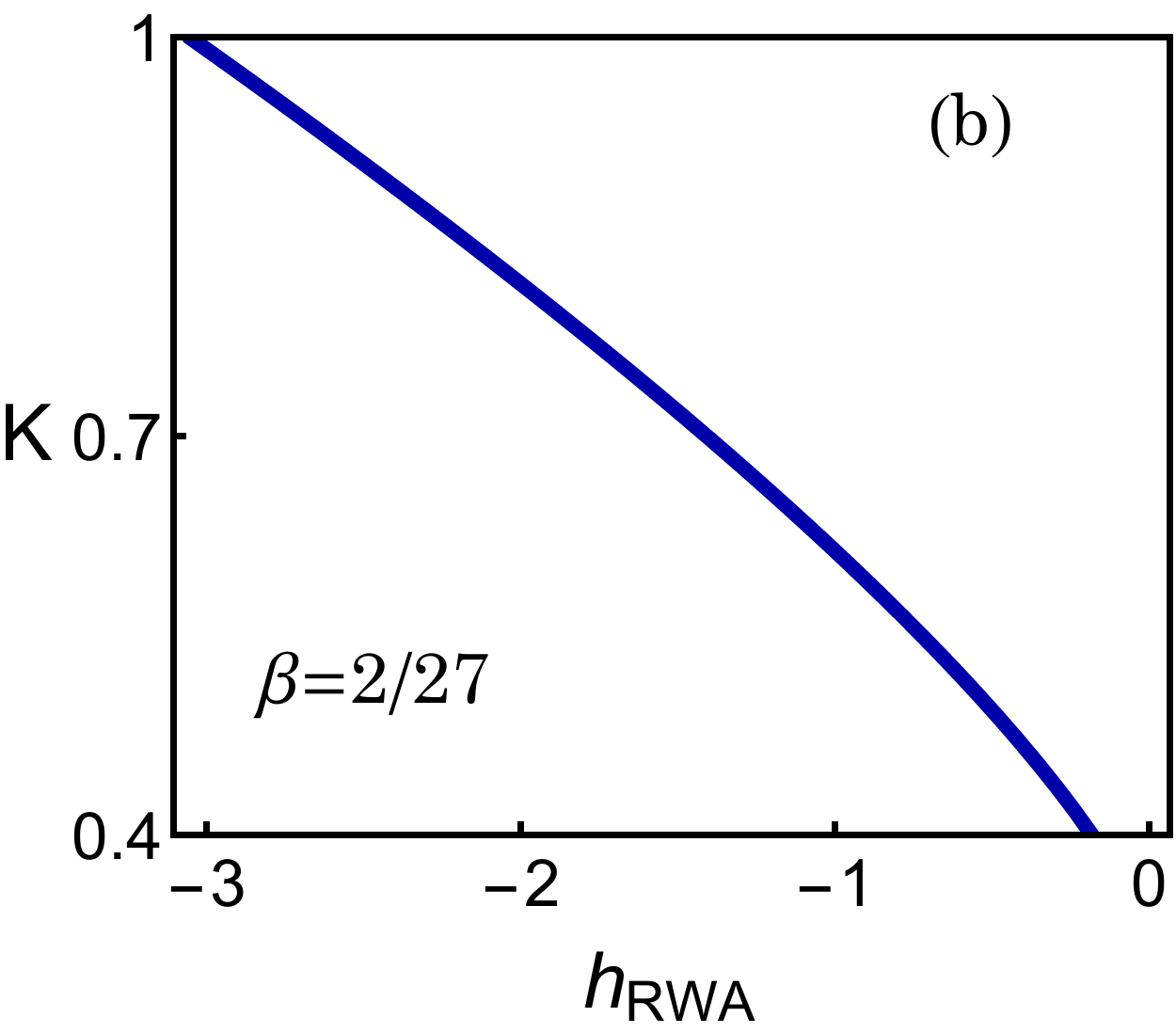}
\caption{ (a) The Hamiltonian trajectories (\ref{eq:RWA_Hamiltonian}) for different values of $h_{\rm RWA}\propto \HR$ and the scaled field strength $\beta=2/27$. The driven oscillator is bistable for this $\beta$, and shown are the trajectories that circle the large-amplitude state at the minimum of $h_{\rm RWA}$, which is stable in the absence of RIFF. The trajectories have a horse-shoe form away from the minimum of $h_{\rm RWA}$ also  where the oscillator is monostable. (b) The scaled ratio of the relaxation rates $K$, Eq.~(\ref{eq:rate_ratio}), as a function of the scaled RWA energy $h\propto \HR$, Eq.~(\ref{eq:scaled_Hamiltonian}).}
\label{fig:rate_ratio}
\end{figure}

For small $\Gamma$ and $J_Tq_{\rm st}$ (the condition is specified below), one can think of the steady motion in the rotating frame as occurring with a constant value of the Hamiltonian $\HR$ along the Hamiltonian trajectory (\ref{eq:RWA_Hamiltonian}), see Fig.~\ref{fig:rate_ratio}(a). This value is determined by  the balance of the damping $\propto \Gamma$ and the RIFF. The dissipative losses $\propto \Gamma$ drive $H_{\rm RWA}$ toward its minimum, whereas the RIFF pumping increases $\HR$.  The stationary value of $\HR$ can be found by averaging over the trajectories (\ref{eq:RWA_Hamiltonian}) the equation of motion for $\HR(q_0,p_0)$, which follows from Eq.~(\ref{eq:eom}). We denote such averaging by an overline,
\[\overline{U(t)} = \frac{1}{{ \mathbb T}(\HR)}\int_t^{t+{\mathbb T(\HR)}}dt'\, U\bigl(t'; \HR\bigr),\]
where $U(t;\HR)$ is a function calculated along the trajectory (\ref{eq:RWA_Hamiltonian}) for a given value of $\HR$ and ${\mathbb T}(\HR)$ is the period of motion along this trajectory. After straightforward algebra we obtain from Eqs.~(\ref{eq:eom})
\begin{align}
\label{eq:balance_H_RWA}
&\overline{d\HR/dt{}} \nonumber\\
&= \frac{1}{{ \mathbb T}(\HR)}\int_{{\cal S}(\HR)}dq_{\rm 0}\,dp_{\rm 0}\left(-2\Gamma +J_Tq_{\rm 0}\right).
\end{align}
Here, ${\cal S}(\HR)$ is the area inside the Hamiltonian trajectory (\ref{eq:RWA_Hamiltonian}) with a given $\HR$.

From Eq.~(\ref{eq:balance_H_RWA}), the stationary (and stable) value of $\HR$ is determined by equation
\begin{align}
\label{eq:rate_ratio}
&\left(J_Tq_{\rm st}/2\Gamma\right)K=1,\nonumber \\
&K=q_{\rm st}^{-1}\int_{{\cal S}(\HR)}q_{\rm 0}\,dq_{\rm 0}\,dp_{\rm 0}\left[\int_{{\cal S}(\HR)}dq_{\rm 0}\,dp_{\rm 0}\right]^{-1}.
\end{align}
Parameter $K$ gives the ratio of the relaxation rates due to the RIFF and the  linear friction. It is the dependence of $K$ on $\HR$ that allows one to find the stable value of $\HR$ from Eq.~(\ref{eq:rate_ratio}). This dependence is illustrated in Fig.~\ref{fig:rate_ratio}(b). 

Figure~\ref{fig:rate_ratio} is plotted in the scaled variables $Q_{\rm 0},P_{\rm 0}$ and for the scaled Hamiltonian $h_{\rm RWA}=(6\gamma/F^4)^{1/3}\HR$, 
\begin{align}
\label{eq:scaled_Hamiltonian}
&h_{\rm RWA}=\frac{1}{4}(Q_{\rm 0}^2+P_{\rm 0}^2)^2 -
 \frac{1}{2}\beta^{-1/3}(Q_{\rm 0}^2 + P_{\rm 0})^2 -Q_{\rm 0}, \nonumber\\
&Q_{\rm 0} = q_{\rm 0}/\zeta, \;P_{\rm 0}=p_{\rm 0}/\zeta, \quad \zeta = (4F/3\gamma)^{1/3}\omega_F^{1/2}.
\end{align}
Function $h_{\rm RWA}$ depends only on one dimensionless parameter, the scaled strength of the driving field 
 \[\beta= 3\gamma F^2/32 \omega_F^3(\delta\omega)^3.\]

As seen from Fig.~\ref{fig:rate_ratio}(b) and also from Eq.~(\ref{eq:scaled_Hamiltonian}), $K=1$ where $h\propto \HR$ is at its minimum. Importantly, $K$ monotonically decreases with increasing $\HR$ in a broad range of $\HR$. This decrease holds both in the range of $\beta$ where the oscillator is bistable and where it is monostable in the absence of the RIFF. Therefore, in the presence of the RIFF, once the condition of the onset of oscillations in the rotating frame is met, $J_T q_{\rm st}>2\Gamma$, these oscillations are stabilized at the value of $\HR$ given by $K\equiv K(\HR) =( 2\Gamma/J_T q_{\rm st}<1$. We emphasize that the frequency of these oscillations $2\pi/{\mathbb T}(\HR)$ is small compared to $\omega_F$, yet it exceeds $\Gamma$ and $J_Tq_{\rm st}$.

The parameter $J_T q_{\rm st}$ depends on the amplitude of the driving field $F$ and the frequency $\omega_F$. By varying $F$ and $\omega_F$ one can control the stable value of $\HR$ and thus the amplitude and frequency of the oscillations in the rotating frame. Remarkably, these oscillations  become significantly nonsinusoidal already for comparatively small difference between $\HR$ and its minimal value. This is seen in Fig.~\ref{fig:rate_ratio}(b). The profoundly non-elliptical trajectories are a signature of nonsinusoidal vibrations. Formally, the oscillations are described by the Jacobi elliptic functions, which allows finding their Fourier components in the explicit form \cite{Dykman1988a}. 

The instability of the forced vibrations at the drive frequency and the onset of nonlinear self-sustained oscillations in the rotating frame lead to a qualitative change of the power spectrum of the driven oscillator. The $\delta$-shaped peak at the drive frequency $\omega_F$ disappears, and instead there emerge  multiple equally spaced peaks on the both sides of $\omega_F$ that correspond to the vibration overtones in the rotating frame. The spacing between the peaks is small compared to $\omega_F$. The widths of the peaks are determined by phase diffusion due to the noise in a nanoresonator, in particular, the thermal fluctuations of $\HR$ around its stable value and the related fluctuations of the frequency $2\pi/{\mathbb T}(\HR)$. These fluctuations are efficiently averaged out by the relaxation, the process reminiscent of motional narrowing in NMR \cite{DK_review84,Maillet2017}. Therefore the widths of the peaks should be much smaller than the damping rate $\Gamma$. Such behavior has indeed been observed in the experiment [20].    

In conclusion, we have shown that, from the symmetry and resonance arguments, a resonantly driven vibrational mode can experience a specific friction force. This force, the RIFF, is nonlinear in the mode coordinate and explicitly depends on the driving force. We considered a microscopic mechanism of the RIFF in nanomechanics associated with the driving-induced spatially nonuniform heating of a nanoresonator and the resulting change of the mode eigenfrequency. The RIFF can be negative. In this case, already for a moderately strong resonant drive, it can lead to an instability of forced vibrations  of a weakly damped nonlinear mode, qualitatively modifying the familiar response of such a mode to a resonant drive. The instability causes the onset of self-sustained oscillations in the rotating frame. In turn, this leads to a characteristic structure of the power spectrum of the driven mode.

MD is grateful for the warm hospitality at Caltech. MD and MLR acknowledge partial support from the NSF (Grant no. DMR-1806473). GR and EMW acknowledge partial support from Deutsche Forschungsgemeinschaft through the collaborative research center SFB 767. 

\appendix

\section{Thermally induced nonlinear friction}
\label{sec:Appendix_thermal}

Here we discuss the temperature change and the resulting change of the vibration eigenfrequency of a resonantly driven nanomechanical resonator. In the units used in the main text, where we set the effective mass equal to unity, the displacement at the mode antinode has dimension $[q] = \rm g^{1/2}  cm$, whereas the resonant force has dimension $[F]=\rm g^{1/2}cm/s^2$. If we consider a flexural mode in a quasi one-dimensional beam or  a string, the displacement as a function of the coordinate $x$ along the beam is $u(x,t) = \rho_{1D}^{-1/2}\phi(x) q(t)$, where $\rho_{1D}$ is the density per unit length and $\phi(x)$ gives the shape of the mode, $\int \phi^2 dx = 1$. The energy in the driving field is $-\int dx f(x,t)u(x,t)$, where $f(x,t)$ is the ``true'' force per unit length. If we think of the force-induced term in the equation of motion  as $[\rho_{1D}\ddot u(x)]_F = f(x,t)$, then we have for the force in the equation for $q$ [Eq.~(3) of the main text] the expression $F(t)=\rho_{1D}^{-1/2}\int dx\,f(x,t)\phi(x)$. Experiments on nanomechanical systems can be usually well described if one assumes that the force $f(x,t)$ can be factored into a space- and time-dependent parts, $f(x,t) = \tilde f_{\rm sp}(x)f_{\rm t}(t)$. Then
\begin{align}
\label{eq:force_F}
F(t) =\rho_{1D}^{-1/2}f_{\rm t}(t)\int dx\, f_{\rm sp}(x)\phi(x).
\end{align}

The power dissipated by the force per unit length is $f(x,t)\partial_t u(x,t)$. For a uniform isotropic resonator, the full equation for the increment of the temperature field is 
\begin{align}
\label{eq:T_conductivity}
C_r\partial_t \delta T = k_r\partial_x^2\delta T +f(x,t)\partial_t u(x,t)/S,
\end{align}
where $C_r$ is the specific heat of the resonator per unit volume, $k_r$ is the thermal conductivity, and $S $ is the cross-section area. We assume here that the temperature is constant across the resonator; an extension to a more general case, including the Zener thermoelastic relaxation \cite{Zener1937} (see also \cite{Lifshitz2000}) is beyond the scope of this paper.

Equation (\ref{eq:T_conductivity}) has to be complemented by the boundary conditions. Often it is assumed that the temperature at the boundary of a nanoresonator is fixed by the support \cite{Chien2018}, a condition that applies if the support has a large mass and a high thermal conductivity, for example. Equation~(\ref{eq:T_conductivity}) can be then solved by expanding $\delta T(x,t)$ in the orthogonal eigenmodes $T_n(x)$ of the temperature field in the absence of the drive,
\[(k_r/C_r)\partial_x^2 T_n = -\lambda_nT_n, \quad \int dx T_n(x)T_m(x)=\delta_{nm}\]
(the analysis can be easily extended to a more complicated geometry of the resonator and to more complicated boundary conditions than $T_n=0$). 

The major contribution to the temperature change  comes from the mode $T_n(x)$ that has the form close to that of $f(x)\phi(x)$. It depends  on the boundary conditions for the temperature field, the spatial structure of the displacement field of the mode $\phi(x)$, and also the coordinate dependence of the driving field. 

For dielectric nanoresonators the thermal conductivity is comparatively low. At room temperature $k_r \sim 10^6~{\rm erg/(cm\cdot s\cdot  K)}$, and the 
specific heat is $C_r\sim 10^7~{\rm erg/(cm^3\cdot K)}$. Then for the resonator length $l_r\sim 10~\mu$m, the relaxation time of low-lying thermal modes is  $\tau_T\sim C_rl_r^2/k_r\sim 10^{-5}$~s. This time significantly exceeds the period (reciprocal frequency) of the vibrational modes, which is typically below $10^{-6}-10^{-7}$~s. Then the temperature field averages out the oscillating terms in $f(x,t)\partial_t u(x,t)$ in Eq.~(\ref{eq:T_conductivity}). At the same time, $\tau_T$ is typically much shorter than the relaxation times of  low-lying vibrational modes, which often exceeds the vibration period by a factor $>10^4$. In this important case the temperature adiabatically follows the vibration amplitude.

The driving-induced temperature change is then of the form $\delta T(x,t)=\sum c_n(t)T_n(x)$ with 
\begin{align}
\label{eq:T_coefficients}
c_n=\rho_{1D}^{-1/2}(SC_r)^{-1}\lambda_n^{-1}\int dx \,T_n(x)\phi(x)[f(x,t)\,\dot q]_{\rm av},
\end{align}
where $[\ldots]_{\rm av}$ indicates averaging over the vibration period. For low-lying vibrational modes and for a weakly nonuniform driving force $f(x,t)$ the major contribution to $\delta T(x,t)$ comes from low-lying temperature modes, with $\lambda_n\sim 1/\tau_T$. Then the magnitude of the temperature change averaged over the resonator is
\[ \delta T \sim l_r^2 k_r^{-1}S^{-1}\, [ F(t)\dot q(t)]_{\rm av}. \] 
We note that the assumption of the temperature being constant in the resonator cross-section requires that $C_rl_\perp^2/k_r$ ($l_\perp$ is the typical transverse dimension) be much shorter than the vibration period, the condition well satisfied for the typical $l_\perp \lesssim 0.1~\mu$m. 

The temperature change causes a change of the vibration frequency. There are several mechanisms of this effect \cite{Atalaya2016}. One of them is the coupling of the mode to the phonons in the nanoresonator that is nonlinear in the mode strain. This coupling is fairly general. It emerges already from the combination of the standard cubic coupling of the considered low-frequency mode (in particular, a flexural mode) to acoustic phonons and the geometric nonlinearity, but it also comes from other terms in the nonlinear Hamiltonian of the vibrations in the resonator. 

Phenomenologically, the mechanism can be described by taking into account the term in the free energy density of the nanoresonator $\delta {\cal F}$, which is quadratic in the linear strain tensor $\hat\epsilon(\rb)$ and linear in the temperature change $\delta T(\rb)$. A simplified form of this term in the one-dimensional model  for a flexural mode is 
\begin{align}
\label{eq:T_dependent_F}
\delta{\cal F} = -\gamma_{\cal F}\int dx \,\delta T(x)(\partial_x^2 u)^2,
\end{align}
where $\gamma_{\cal F}$ is the coupling constant; it is determined by the thermal expansion coefficient, the specific heat, and the resonator geometry \cite{Atalaya2016}. The elastic part of the free energy in the harmonic approximation can be written as ${\cal F}_E = \frac{1}{2}\gamma_\omega \int dx\,[\partial_x^2 u]^2$ with $\gamma_\omega$ determined in the standard way by the elasticity and the geometry \cite{LL_Elasticity}; this term gives the vibration frequency $\omega_0$ for constant temperature. It corresponds to the potential energy of the mode written as $\omega_0^2q^2/2$.

\begin{widetext}
Then the change of the vibration frequency due to the temperature change is 
\begin{align}
\label{eq:freq_change_flexural}
\delta\omega_0 =-(\omega_0\rho_{1D})^{-1}\gamma_{\cal F}\int dx \,\delta T(x) (\partial_x^2\phi)^2.
\end{align}

From Eqs.~(\ref{eq:force_F}), (\ref{eq:T_coefficients}),  and (\ref{eq:freq_change_flexural}) we find that, for a slow thermal relaxation, the resonant driving induced force in the equation for $q(t)$ is
\begin{align} 
\label{eq:force_deriverd}
f_T = &G_T [F(t)\dot q(t)]_{\rm av}q(t), \qquad G_T=2\gamma_{\cal F}(\rho_{1D}SC_r)^{-1}
\sum_n\lambda_n^{-1}\int dx\,T_n(x)\phi(x)f_{\rm sp}(x)\nonumber\\
&\times \int dy T_n(y)(\partial_y^2\phi)^2\left[\int dx f_{\rm sp}(x)\phi(x)\right]^{-1}.
\end{align}
The coefficient $G_T$ gives the coefficient $2m\omega_0\lambda_\omega\lambda_T$ in Eq. (2) of the main text, with the account taken of the spatial dependence of the temperature change.
\end{widetext}

It should be noted that the coupling (\ref{eq:T_dependent_F}) also leads to the standard nonlinear friction, with the friction force that corresponds to $q^2\dot q$ or $\dot q^3$ in the phenomenological picture \cite{Atalaya2016}. However, in the considered case of slow thermal relaxation this force has an extra factor $\propto (\tau_T\omega_0)^{-2}$. Therefore it can be small compared to the force $f_T$.

In prestressed nanoresonators, an important mechanism of the coupling of the frequency and temperature changes is related to the change of the tension due to thermal expansion, cf. \cite{Chien2018} and references therein. It can be analyzed in a way similar to that described above and leads to a qualitatively similar result. If the thermal expansion coefficient is positive, this mechanism leads to the decrease of the vibration frequency with an increasing drive strength, as does the geometric nonlinearity.



%

\end{document}